# Implementation and User Acceptance of Research Information Systems

An Empirical Survey of German Universities and Research Organisations

Otmane Azeroual[1,2], Joachim Schöpfel[3], Gunter Saake[2]

[1] German Center for Higher Education Research and Science Studies (DZHW), Berlin, Germany; E-Mail: *Azeroual@dzhw.eu*
[2] Otto von Guericke University Magdeburg, Institute for Technical and Business Information Systems – Database Research Group, Magdeburg, Germany; E-Mail: *Saake@iti.cs.uni-magdeburg.de*
[3] University of Lille, GERiiCO Laboratory, Villeneuve-d'Ascq, France; E-Mail: *Joachim.schopfel@uni-lille.fr*

## Abstract

**Purpose:** The paper presents empirical evidence on the implementation, acceptance and quality-related aspects of research information systems in academic institutions.
**Design/methodology/approach:** The study is based on a 2018 survey with 160 German universities and research institutions.
**Findings:** The paper presents recent figures about the implementation of RIS in German academic institutions, including results on the satisfaction, perceived usefulness and ease of use. It contains also information about the perceived data quality and the preferred quality management. RIS acceptance can be achieved only if the highest possible quality of the data is to be ensured. For this reason, the impact of data quality on the technology acceptance model is examined, and the relation between the level of data quality and user acceptance of the associated institutional RIS is addressed.
**Research limitations/implications:** The data provide empirical elements for a better understanding of the role of the data quality for the acceptance of RIS, in the framework of a technology acceptance model. The study puts the focus on commercial and open-source solutions while in-house developments have been excluded. Also, mainly because of the small sample size, the data analysis was limited to descriptive statistics.
**Practical implications:** The results are helpful for the management of RIS projects, to increase acceptance and satisfaction with the system, and for the further development of RIS functionalities.
**Originality/value:** The number of empirical studies on the implementation and acceptance of RIS is low, and very few address in this context the question of data quality. Our study tries to fill the gap.

## Keywords

# Introduction

The volume of research activities, structures and staff is steadily growing, and research institutions must be able to answer to increasing demands on research evaluation and reporting by researchers, funders, authorities and the public change. The usual way to deal with this increasing demand is high-performing research information management (RIM). RIM can be defined as "the aggregation, curation, and utilization of metadata about research activities" (Bryant et al. 2017), i.e. processing information about projects, results, structures, persons, infrastructures, equipment, facilities etc., and it involves human resources (HR) as well as organisation and technology in order to produce useful and reliable knowledge about research.

Often, information on research activities is stored in different data bases, such as HR and accounting systems, library catalogues or repositories, and research institutions do their best to get this information out of those data silos and aggregate them in an appropriate and efficient way. A growing number of universities, research organisations and councils implement specific systems in order to improve the RIM quality, to reduce time and efforts and to provide standardized and interoperable knowledge, compliant with requirements by funders and authorities. These research information management systems (RIMS or shortly RIS) are designed to support research institutions in the provision of funding information and reporting, in aggregating references for research outputs, and in producing indicators and assessment (de Castro 2018).

Sometimes a RIS is middleware based on a common exchange format while in other institutions it substitutes existing systems. In any case, implementing a RIS is an important, long-term investment with significant impact of the organisational structure, the information system as a whole and the job profiles. Also, the implementation process requires special attention, and key factors of success must be handled carefully, such as acceptance and quality. A growing body of papers has been published on different aspects of successful implementation of RIS in research institutions. Based on research on data quality in the field of research information management, this paper presents empirical evidence on the implementation, acceptance and quality-related aspects of RIS, with results from a recent survey with 160 German universities and research organisations.

# Literature overview

The euroCRIS repository[1] contains more than 600 conference papers, articles and reports on current research information systems, and several others are referenced by Google

---
[1] https://dspacecris.eurocris.org/

Scholar. However, only few deal directly with user acceptance and data quality as key elements of the implementation process.

## Concept and purpose of research information systems

Following a recent report from OCLC, "research information management systems collect and store metadata on research activities and outputs such as researchers and their affiliations; publications, datasets, and patents; grants and projects; academic service and honors; media reports; and statements of impact" (Bryant et al. 2017). They can be described as specialized databases or specialized federated information systems in which distributed information about research produced from different sources (administration, science…) is aggregated in order to provide a structured view of the equipment and services of an organization and its organizational units (Azeroual et al. 2018a, d).

The nomenclature of research information management practices is to some degree unstandardized and regionalized. The European term of current research information systems (CRIS) is largely unused in North America, "where an alphabet soup of terms are proliferating as the ecosystem matures" (Bryant et al. 2018), such as RNS (Research Networking System), RPS (Research Profiling System) or FAR (Faculty Activity Reporting). In the following, the preferred term is RIS (Research Information System), used as an umbrella concept, independent of the software development model and license, of the business model, of data models and formats, and of the system architecture. Two key elements are shared by all RIS, i.e. the system purpose and the ingested data.

Purpose: In a general way, RIS are relevant for the evaluation, the monitoring and the governance of research activities in a given institution or research organisation. More concretely, RIS are considered helpful for the management of annual reporting, for monitoring and ensuring institutional compliance with open science policies and other requirements, for supporting strategic decision taking and, less, for improving services for researchers and supporting expertise discovery. RIS can support the entire research process of scientific institutions and offer the opportunity for new knowledge to be discovered and generated in universities and research institutions. Through the integration of different data sources, they can help institutions saving time and money and increasing transparency and efficiency of scientific governance. They also provide a database for value-added services, in particular for web applications using business intelligence tools such as reports, for external reporting and representation.

Data: RIS ingest a large variety and volume of research-related information which has been conceptualized and modelled in different ways, the best-known model being the Current European Research Information Format (CERIF) with different levels of entities and relationships[2]. Primary entities are for instance persons, projects and institutions; on other levels, the format defines research output (publications, patents, products etc.) and other aspects, such as awards, funding, expertise, equipment, metrics, citations etc. CERIF also provides a model for the relationships between entities on the same or different levels. Also, the main entities have been subject to standardization, with unique identifiers, dictionaries, taxonomies, templates etc.[3] Beyond the data model, RIS share another data-related problem, i.e. the data inconsistencies and fragmentation because of different data sources and silos and because of the endless rekeying of the same data.

---

[2] https://www.eurocris.org/cerif/main-features-cerif
[3] Cf. for instance the work of CASRAI https://casrai.org/

Through automatic and standardized extraction, processing, analysis and presentation, RIS can contribute to save human and financial resources, to reduce the risk of errors and to guarantee high data quality, providing decision-makers with reliable and consistent data for efficient strategic decisions.

## Implementing research information systems

Even if the first RIS have been developed more than 30 years ago, they are still considered as a "relatively young product category", and many institutions with a RIS either have it for the first time or have moved to their current system from a locally developed or regional system; a recent survey shows how "commercial and open-source platforms are becoming widely implemented across regions, coexisting with a large number of region-specific solutions as well as locally developed systems" (Bryant et al. 2018).

Different studies confirm a landscape in evolution, and although "the level of market penetration by well-established, feature-rich, mature systems, commercial or otherwise, is still low in some regions" (loc. cit.), a steadily increasing number of institutions and organisation have implemented a RIS, or intend to do so. For instance, on the German RIS market proprietary systems (such as Pure from Elsevier, Converis from Clarivate and Elements from Symplectic/Digital Science) coexist with open-source community-based solutions (such as VIVO, FACTScience from QLEO Science GmbH and DSpace-CRIS).

A small number of surveys provide evidence on the reality of implementing RIS in the field of higher education and research.

Ten years ago, the European Science Foundation (ESF) reported an implementation rate of 65% in a small sample of 26 European research organisations, funders and academies (Mugabushaka 2008). Some years later, another European study, the joint EUNIS and euroCRIS survey with 86 respondents in 20 countries assessed a similar situation insofar 69% of the surveyed institutions had implemented a RIS, most of them along with an institutional repositories and often running the same system, either an in-house built or a commercial solution, e.g. Pure or Converis (Ribeiro et al. 2016).

The recent international euroCRIS/OCLC survey with 381 institutions in 44 countries demonstrated the global nature of research information management (RIM) activities and showed that "over half (58%) of the respondents currently have a live system and another 13% are in the process of implementing RIM capacity" while the others are in the procurement process, are exploring, or not considering implementing a RIM system. Those with a RIS are generally satisfied, and 79% would recommend their solution to other institutions (Bryant et al. 2018).

The situation differs depending on the countries concerned.. In the Nordic countries, Norway plays a leading role for the development and implementation of RIS for nearly 20 years now. However a report from the Swedish Lund University concluded ten years ago that in many institutions the operating RIS were not entirely up-to-date, because "they often do not contain enough data to be truly useful" (Rabow 2009).

Pouliquen and Séroussi (2015) have described the situation of RIS in France as "scattered", with many in-house solutions and a lack of interoperability and standardization. Yet, they also observed a beginning cooperation between universities, research organisations, funding agencies and authorities for a higher degree of integration and data sharing.

A quite different situation has been reported for Greece, where many organisations run their own institutional RIS based on the same technology platform developed and maintained by a central operator and compliant with the European standard format CERIF (Karampekios and Androutsopoulou 2016).

For Peru, in a quite different context, Melgar-Sasieta et al. (2018) report a low implementation rate, with only 15% operating RIS (mostly in-house developments), while 54% of the surveyed institutions are implementing or considering implementation of a RIS. They conclude that up to now, in Peruvian universities and institutions "RIM processes are typically carried out without a RIM system".

More generally, de Castro (2018) described the European RIS landscape as volatile and constantly evolving. One main evolution is the convergence and merger of RIS and institutional repositories, as described by de Castro et al. (2014), which may offer a flexible solution especially in regions with a fairly well-established open access repository network but less RIS implementation.

According to the literature, main opportunities and driving forces are international competition between scientific institutions as well as advancing technologies, standards and networking enabling new ways to the management of research information and the promotion of institutional brand and prestige. "Research information management is rapidly growing in importance within the context of a highly globalized and competitive research landscape (…) Concurrent with increasing globalization and competition has been an intensifying preoccupation with 'world class' prestige, university rankings, and indicators of research and education impact" (Bryant et al. 2017). Also, a centralized infrastructure and/or governance, such as in Greece, or strong reporting requirements from the Government or the research funders, as in the UK, seem helpful for the implementation of RIS on the local level of institutions (cf. Jörg et al. 2014).

On the other hand, the great number of institutional stakeholders in research information management, along with the impact (and fear) of change, present significant barriers for the adoption of RIS. Other barriers may be the resources needed to develop, implement and/or operate an integrated RIS, including functional requirements, budget and staff, especially in larger and research-intensive universities and organisations (cf. Rabow 2009).

## Data quality and research information systems

The crucial role of stakeholders and user groups for the successful implementation of a RIS has been highlighted by a German case study (Herwig and Höllrigl 2012), an importance that requires special attention, e.g. a clear definition of the objectives, application scenarios, data model and functional requirements, an early and closely integration of all user groups and stakeholders and an open and proactive communication to all users. But this is not enough, and the same case study insists on the detailed analysis and control of existing data sources, if the structure and quality of the provided data meet the requirements of the RIS.

Often, organizations do not lack data, they lack high quality, analyzable data (Scarisbrick-Hauser and Rouse 2007), i.e. they need data that are complete, timely, valid, consistent and integer (Sebastian-Coleman 2013), data that are fit to use by data consumers (Aljumaili et al. 2016). In the context of RIS, invalid or incorrect data which serve as a basis for decision-making can adversely affect both the strategic orientation of an institution and the acceptance of the RIS.

Good quality data of RIS are a "crucial foundation of any successful monitoring and evaluation strategy" (Mugabushaka and Papazoglou 2012). Data must be trustworthy and "fit for purpose" (Stempfhuber 2008). Without a minimum level of reliability and accuracy of information on persons, organization, projects and results, a RIS will be virtually useless for research management and science policy. Also, as a general rule, the higher the quality of data in RIS and the easier it is to make the reporting and the processed data more accessible and reliable, the greater the acceptance of the users.

Universities and research institutes must clearly define the definition of the type and purpose of data collection as well as their use. For this it is necessary to create and assign responsibilities in institutions and to conduct proactive data quality management. Figure 1 illustrates the relationship between data quality and the stakeholders, i.e. universities and research institutions and users of the RIS.

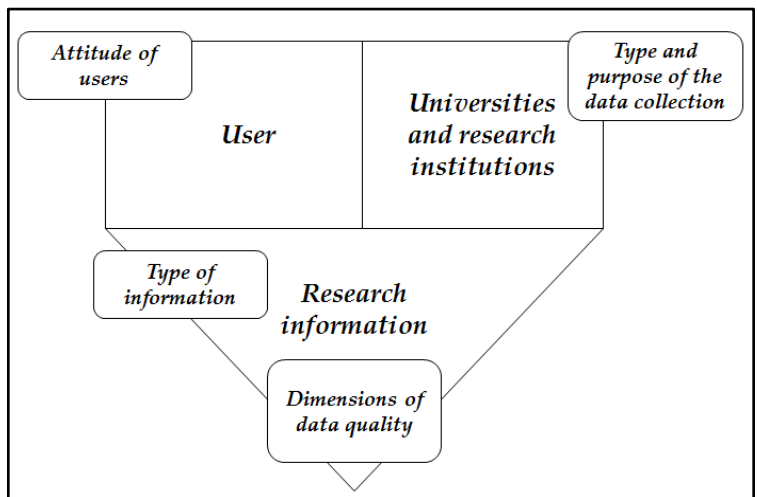

Figure 1: Relationship between data quality and RIS stakeholders

Bad data have a negative effect on decision-making processes and on the implementation of RIS, with the risk of increasing mistrust, wrong decisions in research management and poor acceptance due to the loss of confidence by data users. There can be many reasons for the emergence of quality problems as part of the collection, integration and storage of data sources in the RIS, e.g. data integration of heterogeneous systems (including institutional repositories, cf. Rybinski et al. 2017), structural changes in data sources, a variety of interfaces, the lack of integrity checks, and transformation errors.

To overcome these problems, as a first step, the institutions need to identify the problem that causes quality deficiencies, followed by the need to treat data as an asset in the sense of an internal change in attitude, and finally the implementation of effective quality methods (Azeroual et al. 2018a, b). The successful and sustainable resolution of these data quality issues and the establishment of controls to ensure the quality of ingested data are critical to the success of any of these initiatives.

The euroCRIS/EUNIS study cited above reveals another variable with potential impact on data quality, i.e. the interoperability (linking) with other, internal or external systems (Ribeiro et al. 2016). All reported RIS solutions are linked to other systems, they are not stand-alone systems. The most common links are with HR management systems and institutional repositories, followed by student and financial management systems. Fewer or near to none connections have been reported to other internal systems, like library management, learning management, identity management, organizational management, research equipment databases, data warehouses, appointments, academic partnerships etc.

In contrast, links to external systems are relatively rare, with the connection to research grant management systems being the most frequently reported one, followed by award management and project management systems, which, in other institutions, are available internally. In any case, all these connections to other systems introduce data heterogeneity and increase the risk of inconsistent, invalid or incorrect data.

## Acceptance model and data quality

There are many models for explaining the acceptance of information systems. For the consideration of the acceptance of RIS the most influential technology acceptance model (TAM) is adopted. The TAM was developed by Davis et al. (1989) as an adaptation of Theory of Reasoned Action and originally served to model user adoption of information technologies (IT). The model describes acceptance as the actual use of the technology by its (potential) users on the basis of perceived usefulness (PU) and perceived ease of use (PEOU). The PU refers to a user's assessment of how the use of a specific IT application improves the performance of work tasks within a specific organizational (acceptance) context. The PEOU refers to the assessment of whether the use of the IT system can be learned without difficulty and effort. Thus, the higher the utility and ease of use of a system, the more likely it is for users to use it. In other words, the interaction of both factors results in an intention of the user to use the technology in question (Behavioural Intention to Use, BI). This usage intention may result in actual system utilization of the technology in question (Actual System Use).

The PU and PEOU are valid predictors of the acceptance and actual use of technical systems (Venkatesh and Davis 1996, Arning and Ziefle 2007). Further research on the TAM highlighted the impact of external variables on perceived utility and ease of use (Venkatesh and Davis 2000, Venkatesh and Bala 2008), including data quality as one of the most important conditions for the acceptance. In the context of RIS implementation, data quality includes four aspects (Azeroual et al. 2018c):

- Completeness as the degree to which the system contains all the necessary information (attributes)
- Correctness as perceived accuracy of the information in the system (accordance with reality)
- Consistency as a measure of the perception to what extent the presentation of the information is represented consistently in the system, without contradictions with itself or with other information
- Timeliness as a measure of how well the information is up to date (correspondence with current state of reality)

The perceived data quality is related to the perceived usefulness of the system, as it has a direct impact on the expected benefit of the system use. Figure 2 shows the relationship between data quality and the technology acceptance model.

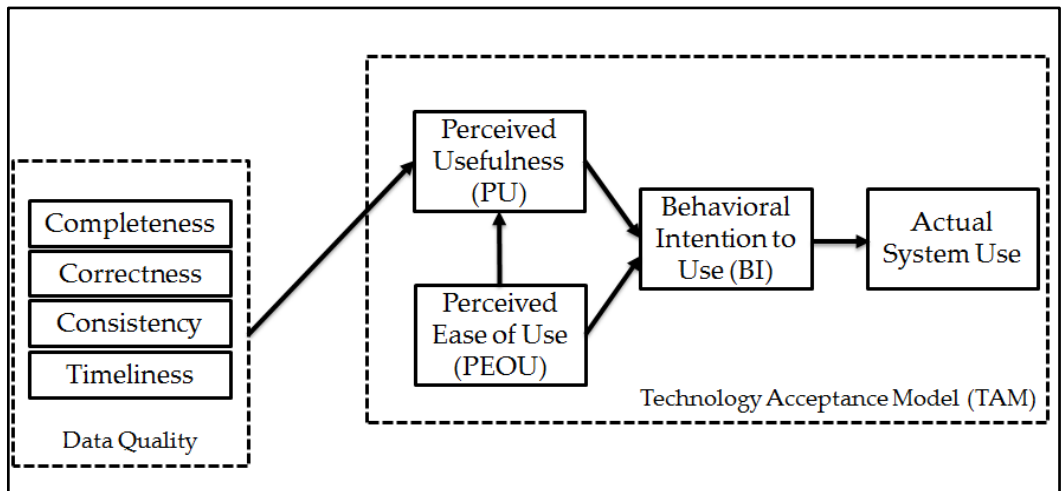

Figure 2: Acceptance model and data quality

The four dimensions of data quality have recently been assessed among users of RIS confirming their reliability and validity (Azeroual et al. 2018c). In particular, the multidimensionality of this data quality construct provides more comprehensive and in-depth information on perceived data quality than the usual surveys and allows targeted data quality management in the field of RIS implementation.

The objective of the following paper is to provide empirical evidence on different aspects of the implementation, the user acceptance and the data quality in the field of research information systems and to contribute to a better understanding and management of acceptance and quality.

# Methodology

In order to assess different aspects of data quality and user acceptance of research information systems, we conducted a survey with an exhaustive sample of 240 German scientific institutions, i.e. 88 universities[4], 54 academic research institutions[5] and 98 public universities of applied sciences[6]. The institutions were invited to respond to an online-based questionnaire[7] with 47 questions:
- 5 questions on the institution and the implementation of a RIS
- 18 questions about data quality and data quality management
- 24 questions about perceived usefulness and ease of use

The survey was carried out on behalf of the *German Centre for Higher Education Research and Science Studies*[8] and took place between May and September 2018.

The contact persons of the sample were mainly administrative staff in the area of controlling, third-party funded projects, research funding & transfer, patents and library, but included also RIS project managers.

---

[4] „Universitäten mit Promotionsrecht" (universities with doctoral programs)
[5] „Universitäre Forschungseinrichtungen"
[6] „Staatliche Fachhochschulen ohne Promotionsrecht" (only Bachelor and Master programs); private universities were excluded
[7] With the *QuestionPro* survey software on https://www.questionpro.de/
[8] Deutsches Zentrum für Hochschul- und Wissenschaftsforschung (DZHW) https://www.dzhw.eu/

The survey was anonymous, and no personal data were processed. All questions were mandatory. 4 questions were open, 12 questions were closed, and 31 questions were four- and five-points rating scales. The results were analysed with descriptive statistics and, for the open questions, with content analysis.

# Results

160 of the 240 contacted institutions responded and completed the online survey, which corresponds to a satisfying response rate of 67%.
51 institutions reported live RIS, such as Pure, Converis, FactScience and VIVO. 36 institutions developed their own local in-house solution. 48 are planning or exploring the future implementation of a RIS (see Figure 3).

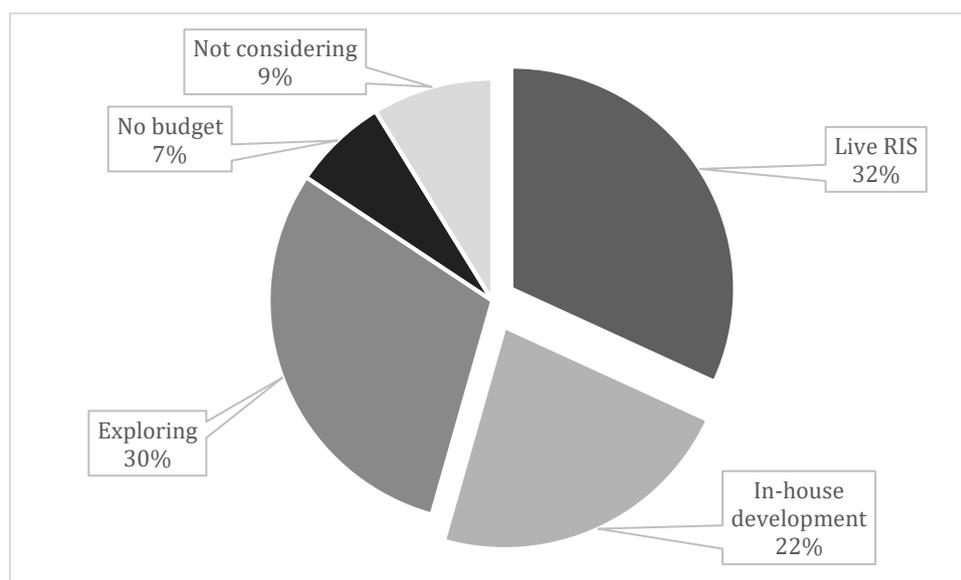

Figure 3. Implementation of RIS (N=160)

Following the results, more than the half of the responding institutions have implemented some kind of a RIS, more often a commercial or open-source solution, less an in-house solution. Together with those institutions which are exploring or preparing to implement a RIS, this means that in the next future nearly all (84%) academic institutions will have their system. Only few institutions report budget problems (7%) or do not consider the use of a RIS (9%). Some of them appear to prefer business intelligence systems, campus management systems (CMS), enterprise resource planning (ERP) or SAP software, project databases like Typo3 or just use simple office applications like Excel.
In the following, the analysis of the results will focus on the responses of the 51 institutions with implemented live RIS, without considering in-house or other non-RIS solutions.

## Perceived data quality and related aspects

The institutional RIS process a large variety of metadata, above all about publications, research staff and third-party funding but also about patents, early career scientists, research infrastructures, events, lectures, awards etc. The responding institutions report an even larger number of internal and external data sources, including identity management systems (IDM), SAP software, ERP systems, project management, financial and human resource management software. Among the external sources, the respondents mention the Web of Science, Scopus, PubMed and the Online Computer Library Center (OCLC) and German National Library catalogues.

A minority of institutions integrate standard identifiers like ORCID and DOI (N=8) or make use of the German Research Core Dataset (RCD) or the European CERIF data model (N=7).

The diversity of data sources may be one reason why only 20 institutions (39%) assess the data quality of their RIS as high while nearly as much, 18 (35%) consider the quality as only average and 8 (16%) even as low.

Questioned about the dimensions of data quality, the respondents rank correctness and completeness as most important aspects of data quality, followed by consistency and timeliness (see Figure 4). In other words, they evaluate above all if the data are correct and complete, and their overall assessment of RIS data quality is mainly based on these two variables.

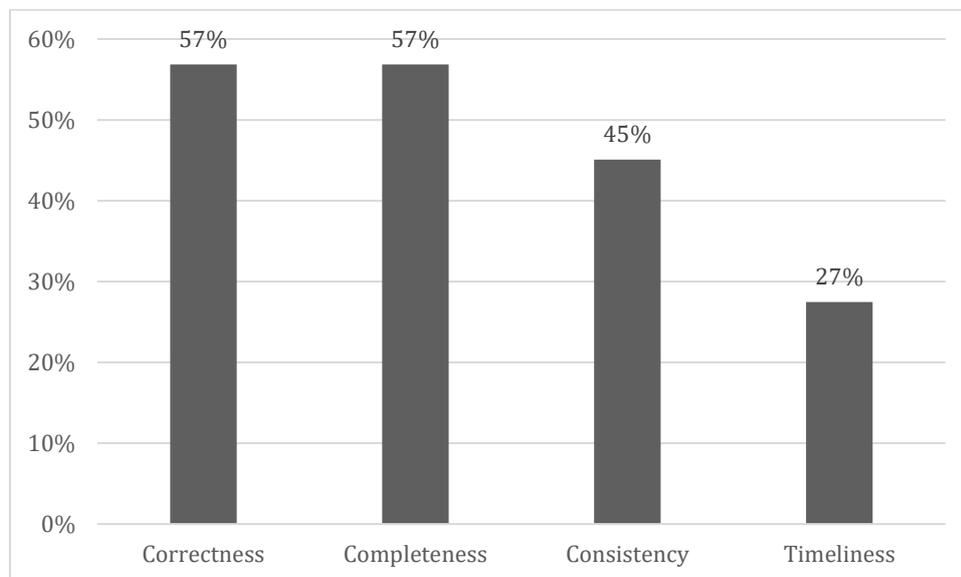

Figure 4. Perceived importance of data quality dimensions (N=51)

How do they deal with data quality issues? Which are their preferred approaches to enhance data quality in their RIS? The respondents rank highest the usual data cleansing methods, including parsing, correction and standardization, enhancement, matching and consolidation (see Figure 5).

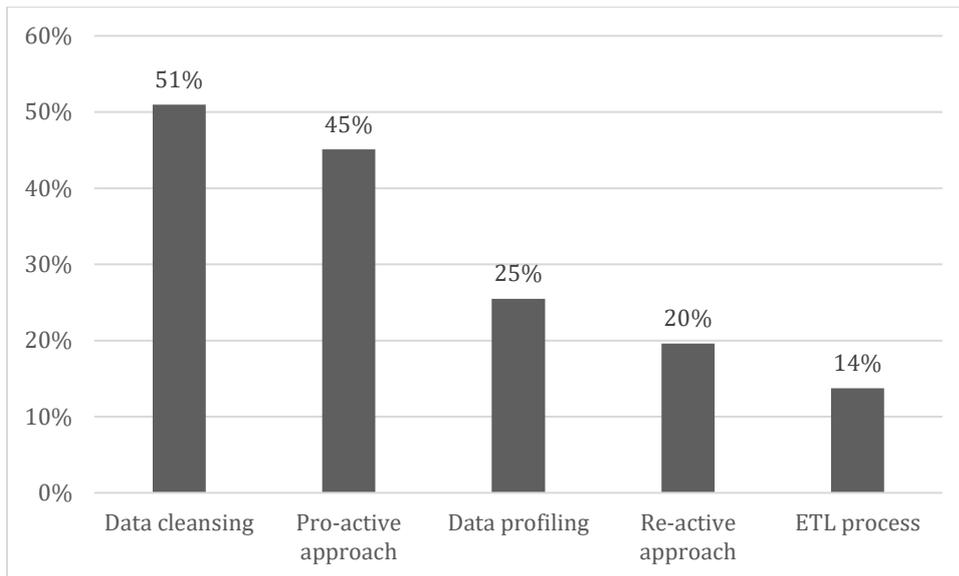

Figure 5. Preferred approaches to data quality management (N=51)

Nearly as important are proactive techniques, available especially for important and frequently changing data. These techniques are designed to eliminate the sources of errors and to prevent the occurrence of the errors, which requires continuous monitoring of possible errors takes place and continuous measures to eliminate and prevent them.
Compared to these two approaches, data profiling, re-active approaches and other techniques to control the filling processes (e.g. extract, transform, load) play a less significant role in the data quality management of RIS, to ensure a high level of correct, complete, consistent and timely data.

## Perceived usefulness

The survey assessed the perceived usefulness of RIS with six five-points rating scales ranging from 5 (strongly agree) to 1 (disagree). The respondents consider their institutional RIS generally as useful, with an average score of 4.3 on the six scales. A very large majority, mostly between 80% and 90%, agrees or strongly agrees that their RIS is useful for their work and that it improves their work performance, productivity and efficiency. For them, the RIS is necessary for their organisation and suits its needs; in other words, it makes sense for their university or research institute. Together and over all six scales, nearly 90% agree (36%) or strongly agree (50%) to the different statements about perceived usefulness. However, the results reveal some particular differences in the response patterns, and their analysis may be helpful for future implementation (see Figure 6).

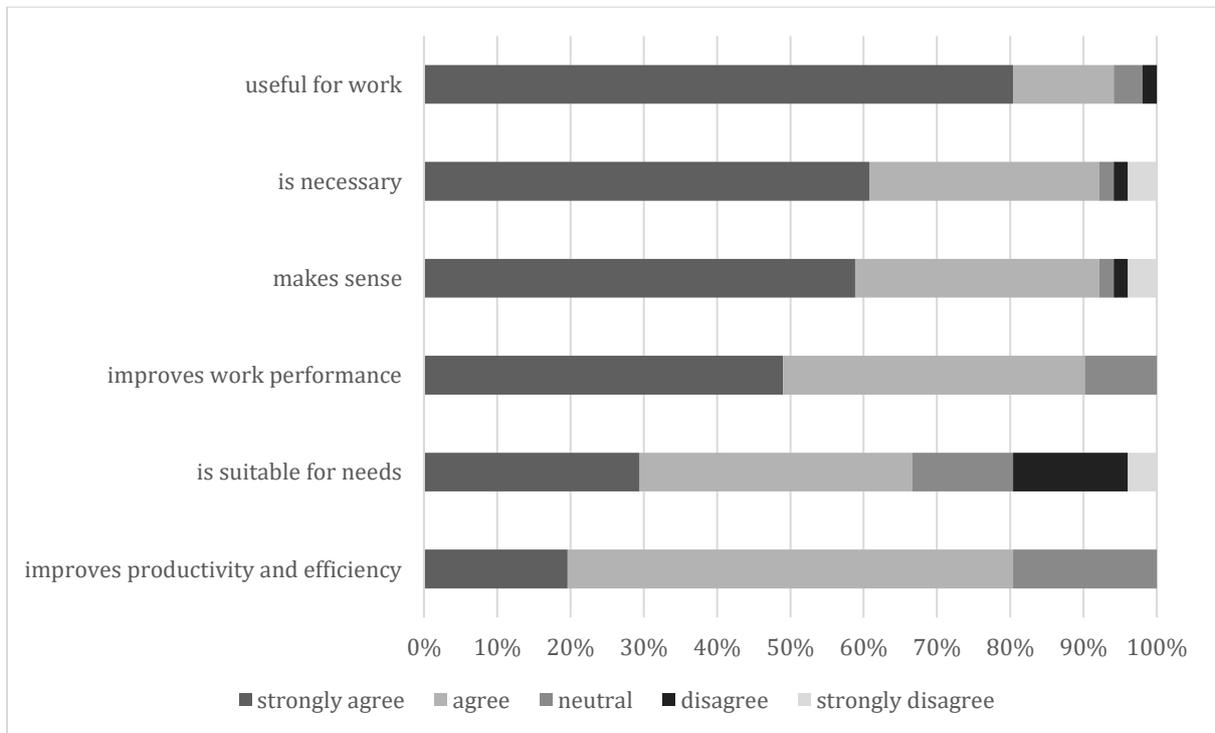

Figure 6. Perceived usefulness of RIS (N=51)

While a large part of the respondents strongly agrees that the implemented RIS is useful and necessary for their institution, their conviction about the system's effect on work performance seems weaker. Only 20% to 30% strongly belief that the RIS suits their needs and that it helps them to increase the work's productivity and efficiency. The average scores on the rating scales are still high, with 3.7 (needs) and 4.0 (productivity) but less than the mean score, and significantly less than the general perception of usefulness.

## Perceived ease of use

Another eight five-points rating scales assessed the perceived ease of use of the RIS. As before, some scales were formulated in a general way (easy to use, easy to learn) while others were more specific, e.g. about helpful assistance, error messages or simple and quick corrections. The average score for these eight scales is 3.4, and lower than for the usefulness scales (4.3). Over all eight scales, only 42% agree (22%) or strongly agree (20%) to the different statements about perceived ease of use. Many ratings are neutral (37%) or even negative, with 19% expressing their dissent with the statements about mental effort, need for external training or ease of learning. Again, the results reveal differences in the response patterns (see Figure 7).

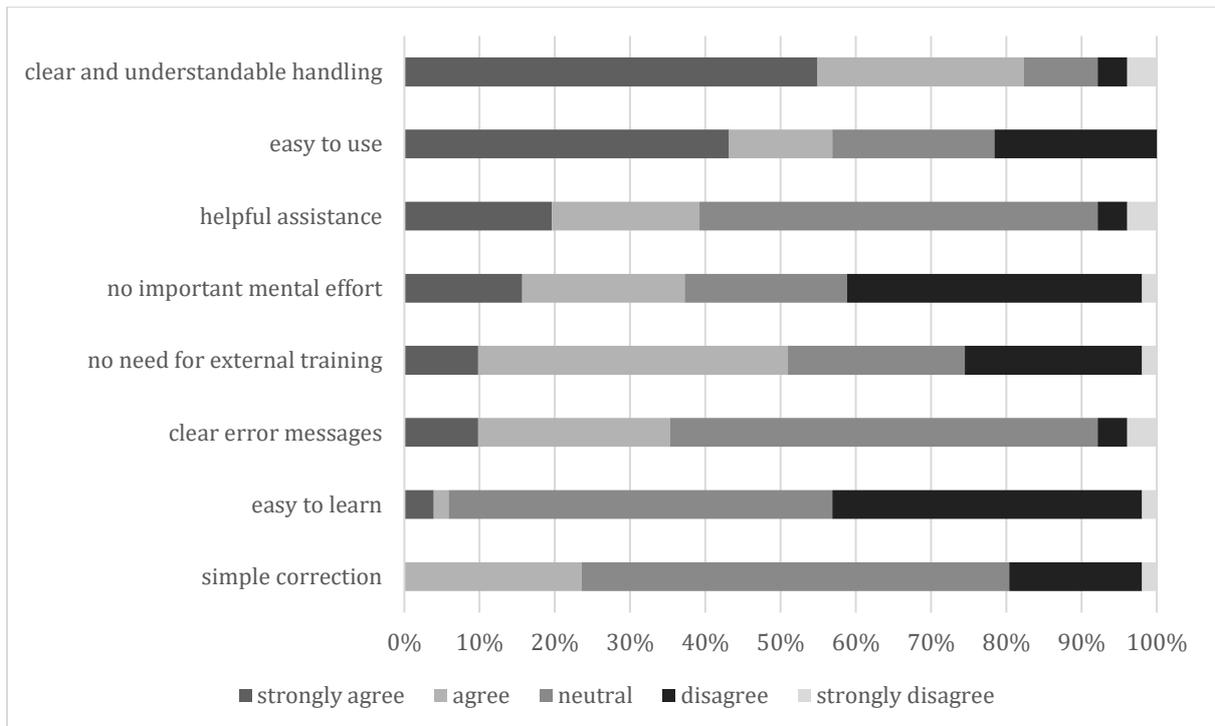

Figure 7. Perceived ease of use (N=51)

Most respondents think that the RIS is more or less easy to understand and to handle, and half of them appreciate that they don't need specific training or even manuals or user guide. On the other side, more than 50% of the respondents feel that the RIS is not easy to learn and that working with the RIS requires significant mental efforts. The system assistance is mainly considered as not very helpful. Many respondents do not describe the RIS error messages as comprehensible or clear, and when the system detects data or processing errors, the required corrective measures (troubleshooting) are generally not appreciated as quick and simple.

Obviously, the survey not only reveals a gap between perceived usefulness and ease of use but also, regarding the latter, a gap between general positive opinions about the ease of use and a more critical stance towards the required mental investment and time to learn and work with the system, especially for the control and correction of data errors. The survey reveals, too, two distinctive groups of respondents, one that is generally satisfied with the system, and another which is less satisfied with the system's functionalities and less convinced of its usefulness. Even so, the overall opinion of the implemented RIS is positive, insofar 86% declare that they would like to recommend their system to other institutions.

# Discussion

## Limitations

The sample can be considered as representative, with a response rate of 67% from an exhaustive sample of German universities and academic research organisations. However, the real number of respondents (51) is rather small, and any analysis must be cautious not to over-interpret the results. For this reason we preferred descriptive statistics to a more complex approach such as predictive analytics. Insofar our study has a more explorative and illustrative character and cannot provide statistical support for the described model of data quality and acceptance.

A second limitation is that the study puts the focus on commercial or open-source community-based RIS solutions and does not consider local in-house developments. The reason of this choice is to obtain a more homogeneous analysis of RIS related aspects. Yet, this approach potentially excludes alternative solutions with a larger range of system architectures and functionalities, different ways of quality management and probably different perceptions by users and administrators.

## Implementation

More than half of the surveyed German universities and academic research institutions have a live RIS. The implementation rate of 54% is similar to other surveys, in particular from the European Science Foundation with European research organizations, funders etc. in 2008 (65%), from the joint EUNIS-euroCRIS survey in European 20 countries in 2016 (69%) and from the global OCLC survey with nearly 400 institutions around the world in 2018 (58%).

The implementation rate of 54% must be seen in the particular context of the German Higher Education and research landscape and as an indicator of a dynamic situation. Germany has no centralized RIS infrastructure or platform, and each institution has to make a decision about how to manage, monitor and report its research activities, with a commercial or open-source full RIS, with a local in-house solution similar to a RIS, or with other tools providing comparable functionalities. There is no centralized incentive or national solution for German institutions, which would probably foster the local implementation of RIS.

On the other hand, the situation is constantly evolving, and about a third of the surveyed institutions declared that they are exploring or preparing a local solution for their research management. The part of institutions responding that they were not interested in a RIS is rather small. This means that in three to five years, depending on the duration of the development and/or implementation process, the situation probably will have changed and significantly more institutions – up to 70-80% - will have a RIS.

Based on the survey and in comparison with other countries, three predictions can be made for the development of the RIS landscape in Germany during the next years:

- As the part of commercial solutions is rather low (32%) and even if the number of potential customers is limited, this part will probably increase, with products from Elsevier (Pure), Clarivate (Converis) and other companies, especially for research-intensive universities.
- Probably, one part of the future RIS will be some kind of "merger solutions", i.e. platforms with functionalities similar to institutional repositories and to research information systems. In other words, in some years some institutions at least will consider their institutional repository as a RIS, either as a local in-house development or with a commercial solution like Pure.
- Finally, an increasing part of the RIS will probably move from the institutional servers on the campus into the cloud, as an infrastructure-, a platform- or a software-as-a-service solution, similar to other parts of the campus information system infrastructure.

## Quality

Data have been considered as an asset (Aljumaili et al. 2016), as "lifeblood" of today's organizations, and data quality is crucial. Lasting improvements in the quality of data require organizational commitment (Sebastian-Coleman 2013). The results show that the German institutions are strongly committed to data quality, especially to the correctness and completeness of data, and they generally prefer data cleansing and proactive approaches to other methods of data quality management. Yet, their overall assessment of data quality is not really satisfying, as about half of the respondents consider their RIS data quality as average or low.

This needs special attention, from at least two sides. On the one hand, RIS managers must be aware of the (lack of) quality of ingested data, i.e. of problems produced outside of the system and their potential negative impact on perceived usefulness and the user acceptance. This means that a thorough assessment of the data sources upstream of the RIS is needed, of their reliability, consistency, formats, standards etc., and that the implementation and interconnection of a RIS must be accompanied by a revision of the data-providing systems, such as repositories, catalogues, HR systems etc. Partial merging of these systems may be an option, like for instance between repositories and RIS (cf. de Castro et al. 2014).

On the other hand, editors of RIS software must take care of data quality management and provide different, reliable and efficient tools for the control and correction of data issues. Just leave the issue to the end user or consider ingested data by default as correct and complete is not realistic. The by default situation is incomplete, incorrect, inconsistent and untimely information, with missing identifiers and unreliable metadata, and the RIS must be able to deal with this situation in a general, data-centered framework (Azeroual et al. 2018a, b, c).

## Satisfaction and acceptance

In spite of the unsatisfying assessment of data quality, most respondents (86%) would like to recommend their system to other institutions, a level which is similar to the 79% of the OCLC survey (Bryant et al. 2018) and which can be interpreted as an indicator for a rather high degree of satisfaction with the RIS. Mediocre data quality and high satisfaction with the system seem paradoxical. However, this result can be compared with the OCLC survey which noted that although "the level of market penetration by well-established, feature-rich, mature systems, commercial or otherwise, is still low in some regions, (…) satisfaction with the products once implemented is often high" (loc.cit.).

Obviously, the relation between perceived data quality and satisfaction is not linear. The technology acceptance model cited above (see Figure 2) and its application to the RIS implementation process may provide elements for a better understanding. As the model shows, the perceived data quality does not impact directly the system acceptance but is moderated by the system's perceived usefulness and ease of use. This means that perceived average or low data quality may not reduce the overall satisfaction with the system, if the system is nevertheless perceived as useful for the institution and the individual user. Indeed, most of the respondents consider their RIS as useful and necessary for their institution and for the improvement of the performance at work, and this is probably the reason for their expressed willingness to recommend the system to other institutions.

Even so, three aspects needs attention.

- Data quality is a multilevel concept, both in terms of the different dimensions of data quality (completeness etc.), as well as the methods of control and corrections of data issues. Both are connected, and both are evaluated and ranked by the respondents differently – some are more important, others are less. Data quality is not a simple concept, and even if some features and/or methods may be evaluated as less satisfying, this assessment does not necessarily impact the overall satisfaction.
- The second aspect is the difference between perceived usefulness and ease of use. The survey shows that one part of the respondents are not really satisfied with the systems' features. They consider the system as rather difficult to learn, and they are not satisfied with the error messages and the needed effort to fix the problems, i.e. the data errors. But again, apparently these problems with initial training and understanding and with everyday troubleshooting do not impact the overall satisfaction with the system which in fact seems more conditioned by the perceived system's usefulness.
- A last aspect should be mentioned. The results indicate that there may be two different groups of respondents, one consistently satisfied with the system and its usage and convinced by its usefulness, the other more critical especially with the system's ease of use. Yet, another interpretation of the results may be a divergent appreciation of the usefulness for the institution and the experienced problems with the system in the everyday work, i.e. a dissonance between the institutional and the individual level. In fact, a system can be assessed as very useful for the institution but as less satisfying for the individual user.

Because of the small number, our data are not representative enough to evaluate these aspects in a more detailed way. Also, they should be addressed in further research, because of their interest for future development and implementation of these systems.

## Conclusion

The reported survey shows that about half of the German universities and academic research institutions have implemented research information systems (RIS) and that they generally consider their RIS as meaningful and useful. While only half of the respondents (49%) assess the RIS data quality as high or good, and even if the opinions on the systems' ease of use are variable, a large majority of 86% would recommend their system to other institutions.

The survey reflects also a dynamic landscape, with a significant part of institutions exploring or preparing the future implementation of a RIS solution. Based on our survey and similar studies from EUNIS, OCLC and euroCRIS, and considering the rapidly changing environment of research management and information technology, special attention should be paid to data quality issues and related features, especially, automatic control and correction, error messages and troubleshooting functionalities, in order to increase the user acceptance and satisfaction with the system. A study with a larger sample of institutions and people (system users and administrators) could be helpful to assess the statistical relationships between the different variables, as indicated by the model.

More generally, two aspects should be addressed as they may impact the data quality issue in a significant way: the merger of the RIS with other systems, especially with institutional repositories, in order to avoid problems with standards, transfers, inconsistency etc.; and the outsourcing of the RIS in the cloud, as an external service, platform or infrastructure.

Additionally, it may be interesting to compare these results with the experience of those institutions which opted for local in-house developments, specific solutions different from the commercial and community-based open-source systems. Does in-house development affect data quality, acceptance and satisfaction? If so, how and why?

# Acknowledgements

The authors wish to express their thanks to the 160 institutions which responded to the survey. The study was funded by the German Centre for Higher Education Research and Science Studies (DZHW) and by the German Federal Ministry of Education and Research (BMBF) as part of the project "Research Core Dataset"[9]. One part of the analysis received funding from the European Institute of Social Sciences and Humanities (MESHS Lille) and from the Regional Council (Conseil Régional Hauts-de-France), as part of the research project "D4Humanities"[10].

---